\documentclass[journal,onecolumn]{IEEEtran}
\usepackage{amsmath}
\usepackage{amssymb}
\usepackage{amsfonts}
\usepackage{graphicx}
\usepackage{epsfig}
\usepackage{subfigure}
\usepackage{psfrag}
\usepackage{xcolor}
\usepackage{amsfonts, bm}
\usepackage{epstopdf}

\makeatletter
\long\def\@makecaption#1#2{\ifx\@captype\@IEEEtablestring%
\footnotesize\begin{center}{\normalfont\footnotesize #1}\\
{\normalfont\footnotesize\scshape #2}\end{center}%
\@IEEEtablecaptionsepspace
\else
\@IEEEfigurecaptionsepspace
\setbox\@tempboxa\hbox{\normalfont\footnotesize {#1.}~~ #2}%
\ifdim \wd\@tempboxa >\hsize%
\setbox\@tempboxa\hbox{\normalfont\footnotesize {#1.}~~ }%
\parbox[t]{\hsize}{\normalfont\footnotesize \noindent\unhbox\@tempboxa#2}%
\else
\hbox to\hsize{\normalfont\footnotesize\hfil\box\@tempboxa\hfil}\fi\fi}
\makeatother

\title{\huge Efficient Reduced-Rank DOA Estimation Algorithms Using Alternating Low-Rank Decompositions}
\author{Yunlong Cai, Linzheng Qiu, Rodrigo C. de Lamare, and Minjian Zhao
\thanks{Y. Cai, L. Qiu and M. Zhao are with the Department of Information Science and Electronic Engineering, Zhejiang University, Hangzhou, China, 310027.
R. C. de Lamare is with CETUC, PUC-Rio Rua Marques de São Vicente, 22493-900
Rio de Janeiro, Brazil and University of York, Heslington, YO10 5DD, York,
England, United Kingdom.
Email: ylcai@zju.edu.cn, lzqiu@zju.edu.cn,  rcdl500@ohm.york.ac.uk, mjzhao@zju.edu.cn.}
\thanks{This work was supported in part by the National Natural Science Foundation of China under Grants 61471319, Zhejiang Provincial Natural Science Foundation of China under Grant LY14F010013, the Fundamental Research Funds for the Central Universities, and the National High Technology Research and Development Program (863 Program) of China under Grant 2014AA01A707.}
}

\begin{document}
\maketitle 

\begin{abstract}
In this work, we propose an alternating low-rank decomposition (ALRD) approach
and novel subspace algorithms for direction-of-arrival (DOA) estimation. In the
ALRD scheme, the decomposition matrix for rank reduction is composed of a set
of basis vectors. A low-rank auxiliary parameter vector is then employed to
compute the output power spectrum. Alternating optimization strategies based on
recursive least squares (RLS), denoted as ALRD-RLS and modified ALRD-RLS
(MARLD-RLS), are devised to compute the basis vectors and the auxiliary
parameter vector. Simulations for large sensor arrays with both uncorrelated
and correlated sources are presented, showing that the proposed algorithms are
superior to existing techniques.
\end{abstract}


\section{Introduction}
Array signal processing has been widely used in areas such as radar, sonar and
wireless communications. Many applications related to array signal processing
require the estimation of the direction-of-arrival (DOA) of the sources
impinging on a sensor array \cite{526899}. Among the well-known DOA estimation
schemes are the Capon method and subspace-based algorithms
\cite{Optimum_array_processing} such as Multiple-Signal Classification (MUSIC)
\cite{1143830} and Estimation of Signal Parameters via Rotational Invariance
Techniques (ESPRIT)\cite{32276}. The Capon method calculates the output power
spectrum for each scanning angle according to the constrained minimum variance
(CMV) criterion. Then the estimated DOAs can be obtained by finding the peaks
of the output power spectrum \cite{1449208}. MUSIC, ESPRIT and their improved
versions \cite{80927,6422415,6680627,4400832,1395953,1341617} estimate the DOAs
by exploiting the signal and the noise subspaces of the signal correlation
matrix. Due to the eigenvalue decomposition (EVD) and/or the singular-value
decomposition (SVD), MUSIC and ESPRIT require a high computational cost,
especially for large sensor arrays. The recently proposed subspace-based
auxiliary vector (AV) \cite{4063549}, the conjugate gradient (CG) \cite{CG} and
the joint iterative optimization (JIO) algorithms \cite{6978860} employ basis
vectors to build the signal subspace instead of the EVD or the SVD. However,
the iterative construction of the basis vectors yields a complexity comparable
to the EVD. Moreover, the AV and CG algorithms cannot provide a satisfactory
performance for large sensor arrays with many sources.

In recent years, large sensor arrays have gained importance for applications
such as radar and future communication systems. The performance of direction
finding algorithms depends on the data record and the array size. Resorting to
large arrays or more snapshots leads to higher resolution
\cite{Optimum_array_processing}. However, direction finding for large arrays
also requires large data records and are associated with high computational
costs. Beamspace DOA estimation \cite{60074,193151,275607,beam_cg_av} is an
effective method to reduce the computational burden. Nevertheless, the
beamspace-based algorithms are sensitive to the presence of sources located
outside the angular sectors-of-interest \cite{1621390}.

In this paper, we present an alternating low-rank decomposition (ALRD) approach
for DOA estimation in large sensor arrays with a large number of sources. In
the ALRD scheme, a subspace decomposition matrix which consists of a set of
basis vectors and an auxiliary parameter vector are employed to compute the
output power spectrum for each scanning angle. In order to avoid matrix
inversions, we develop recursive least squares (RLS) type algorithms
\cite{Adaptive_Filter_Theory} to compute the basis vectors and the auxiliary
parameter vector, which reduces the computational complexity. The proposed DOA
estimation algorithms are referred to as ALRD-RLS and modified ALRD-RLS
(MALRD-RLS), which employs a single basis vector.

The paper is organized as follows. In Section \ref{sec:system_model}, we
outline the system model and the problem of DOA estimation. The proposed ALRD
scheme and algorithms are presented in Section \ref{sec:MALRD}. In Section
\ref{sec:simulation}, we illustrate and discuss the simulation results.
Finally, Section \ref{sec:conclusion} concludes this work.

\section{System Model and Problem Formulation}
\label{sec:system_model}
We consider a uniform linear array (ULA) with $M$ omnidirectional sensor
elements and suppose that $K$ narrowband source signals impinge on the ULA from
directions $\theta_{1}$, $\theta_{2}$, ..., $\theta_{K}$, respectively, where
$M$ is a large number with $K \ll M$. The $i$th snapshot of the received signal
can be expressed by an $M\times 1$ vector as
\begin{equation}
\setlength{\abovedisplayskip}{1pt}
\setlength{\belowdisplayskip}{1pt}
\mathbf{r}(i)= \sum_{k=1}^{K} \mathbf{a}(\theta_{k})s_{k}(i)+\mathbf{n}(i)
\label{eq:system_model}
\end{equation}
where $s_{k}(i)$ is the $k$th source signal with power $\sigma_{s}^{2}$.
$\mathbf{n}(i)$ denotes the noise vector which is assumed to be temporally and
spatially white Gaussian with zero mean and variance $\sigma_{n}^{2}$. The
array steering vector $\mathbf{a}(\theta_{k})$ is defined as
\begin{equation}
\setlength{\abovedisplayskip}{5pt}
\setlength{\belowdisplayskip}{5pt}
{\bf{a}}(\theta_{k})=
[1,e^{-2\pi j\frac{d_{s}}{\lambda_{c}}\cos\theta_{k}},...,
e^{-2\pi j(M-1)\frac{d_{s}}{\lambda_{c}}\cos\theta_{k}}]^{T}
\label{eq:steering_vector}
\end{equation}
where $(\cdot)^{T}$ denotes the transpose operation and $\lambda_{c}$ is the
signal wavelength. The parameter $d_{s}=\frac{\lambda_{c}}{2}$ represents the
array inter-element spacing.

Direction finding algorithms aim to estimate the
DOAs $\bm{\theta}=[\theta_{1},\ldots,\theta_{K}]^{T}$ by processing
$\mathbf{r}(i)$. The correlation matrix of $\mathbf{r}(i)$ is given by
\begin{equation}
\setlength{\abovedisplayskip}{1pt}
\setlength{\belowdisplayskip}{1pt}
\mathbf{R}=\mathbb{E}\{\mathbf{r}(i)\mathbf{r}^{H}(i)\}=
\sum_{k=1}^{K}\mathbf{a}(\theta_{k})\mathbf{R}_{s,k}\mathbf{a}^{H}(\theta_{k})+\sigma_{n}^{2}\mathbf{I}_{M}
\label{eq:autocorrelation}
\end{equation}
where $\mathbb{E}\{\cdot\}$ denotes expectation, $(\cdot)^{H}$ is the Hermitian
operator and $\mathbf{I}_{M}$ is the identity matrix with dimension $M$.
$\mathbf{R}_{s,k}$ is the correlation matrix of the $k$th signal with
$\mathbf{R}_{s,k}=\mathbb{E}\{s_{k}(i)s_{k}^{H}(i)\}$.
$\mathbf{R}_{n}=\mathbb{E}\{\mathbf{n}(i)\mathbf{n}^{H}(i)\}=\sigma_{n}^{2}\mathbf{I}_{M}$
is the correlation matrix of the noise vector. Note that the exact knowledge of
$\mathbf{R}$ is difficult to obtain, thus estimation by sample averages is
employed in practice, which is
$\widehat{\mathbf{R}}=\frac{1}{N}\sum_{i=1}^{N}\mathbf{r}(i)\mathbf{r}^{H}(i)$,
with $N$ being the number of available snapshots.

\section{Proposed ARLD Scheme ALRD-RLS and MALRD-RLS Algorithms}
\label{sec:MALRD} In this section, we detail the proposed ALRD
scheme and the ALRD-RLS and MALRD-RLS DOA estimation algorithms. The
ALRD scheme divides the received vector into several segments and
processes each segment with an individual basis vector. The basis
vectors constitute the columns of the decomposition matrix, which
performs dimensionality reduction
\cite{int,jiolms,jiomimo,jiols,jidf,fa10,saabf,barc}. Then, a lower
dimensional data vector is processed by the auxiliary parameter
vector to construct the output power spectrum. The ARLD-RLS and the
MARLD-RLS algorithms are based on an alternating optimization
procedure of the basis vectors and the reduced-rank auxiliary
parameter vector. \vspace{-0.5em}
\begin{figure}[h]
\centering\scalebox{0.7}{\includegraphics{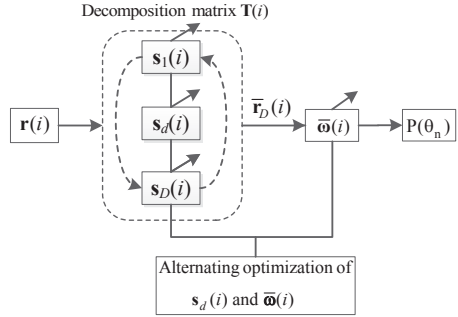}} \vspace{-1em}
\caption{Block diagram of the ALRD scheme} \label{fig:MALRD_block}
\end{figure}
\vspace{-0.2em}

\subsection{Proposed ALRD Scheme}

\label{ssec:MALRD_overview} The block diagram of the ALRD scheme is
depicted in Fig. \ref{fig:MALRD_block}. The received vector
$\mathbf{r}(i)=[r_{0}(i)\ldots r_{M-1}(i)]^{T}$ is processed by an
$M \times D$ decomposition matrix ${\bf T}(i)$ which consists of a
set of $I \times 1$ basis vectors $\mathbf{s}_{d}(i)$, where $d\in
\{1,\ldots D\}$. The resulting data vector can be expressed by
\begin{equation}
\setlength{\abovedisplayskip}{3pt} \setlength{\belowdisplayskip}{3pt}
{\bf{\bar{r}}}_{D}(i)= {\bf T}^H(i){\bf r}(i)
=\sum_{d=1}^{D}{\bf{q}}_{d}{\bf{d}}^{H}_{d}\bm{\mathcal{R}}(i){\bf{s}}_{d}^{*}(i)
\end{equation}
where ${\bf{q}}_{d}$ is a $D\times1$ vector with a one in the $d$th position
and zeros elsewhere. ${\bf{d}}_{d}$ is the $D\times1$ selection vector to
divide ${\bf{r}}(i)$ into $D$ segments, which are defined as:
\begin{equation}
\setlength{\abovedisplayskip}{3pt}
\setlength{\belowdisplayskip}{3pt}
{\bf{d}}_{d}=[~\underbrace{0~,\ldots~,0~}_{\mu_{d}~  zeros},~1~,
\underbrace{0~,\ldots~,0~}_{M-\mu_{d}-1~zeros} ]^{T}
\end{equation}
where $\mu_{d}$ is the selection pattern which is chosen as
$\mu_{d}=(d-1)\lfloor \frac{M}{D} \rfloor$. The $M\times I$ matrix
$\bm{\mathcal{R}}(i)$ corresponding to $\mathbf{r}(i)$ has a Hankel structure
\cite{Matrix_Computations}, which is described by
\begin{equation}
\bm{\mathcal{R}}(i)=
\left(
    \begin{smallmatrix}
    r_{0}(i) & r_{1}(i) & \ldots & r_{I-1}(i) \\
    \vdots & \vdots & \ldots & \vdots \\
    r_{M-I}(i) & r_{M-I+1}(i) & \ldots & r_{M-1}(i) \\
    r_{M-I+1}(i) & r_{M-I+2}(i) & \ldots & 0 \\
    \vdots & \vdots & \ddots & \vdots \\
    r_{M-2}(i) & r_{M-1}(i) & 0 & 0\\
    r_{M-1}(i) & 0 & 0 & 0 \\
  \end{smallmatrix}
\right)
\label{eq:Ro_matrix}
\end{equation}
The $I \times 1$ basis vectors $\mathbf{s}_{1}(i) \ldots \mathbf{s}_{D}(i)$
form the $M\times D$ decomposition matrix $\mathbf{T}(i)$. After dimensionality
reduction, the $D \times 1$ data vector ${\bf{\bar{r}}}_{D}(i)$ is processed by
an auxiliary parameter vector ${\bm{\bar{\omega}}}(i)$ to compute the output
power spectrum. As seen from Fig. \ref{fig:MALRD_block}, the basis vectors
$\mathbf{s}_{d}(i)$ and the auxiliary parameter vector ${\bm{\bar{\omega}}}(i)$
are alternately optimized according to some prescribed criterion, which is
introduced in what follows.

\vspace{-0.2em}
\subsection{Proposed ALRD-RLS DOA Estimation Algorithm}
\label{ssec:RLS-MALRD}

The ALRD-RLS algorithm solves the optimization problem:
\begin{equation}
\setlength{\abovedisplayskip}{2pt}
\setlength{\belowdisplayskip}{2pt}
\begin{split}
& \min_{\bm{\bar{\omega}}(i),\mathbf{s}_{d}(i)} \sum_{l=1}^{i}\alpha^{i-l}
\big|\bm{\bar{\omega}}^{H}(i)\sum_{d=1}^{D}{\bf{q}}_{d}{\mathbf{d}}^{H}_{d}\bm{\mathcal{R}}(l){\bf{s}}_{d}^{*}(i)\big|^{2}\\
& \textrm{subject to} \quad
\bm{\bar{\omega}}^{H}(i)\sum_{d=1}^{D}{\bf{q}}_{d}{\mathbf{d}}^{H}_{d}\bm{\mathcal{A}}_{n}{\bf{s}}_{d}^{*}(i)=1
\end{split}
\label{eq:cost_function}
\end{equation}
where $\alpha$ is a forgetting factor close to but smaller than $1$.
$\bm{\mathcal{A}}_{n}$ is the $M\times I$ Hankel matrix of the scanning
steering vector ${\bf{a}}(\theta_{n})=[a_{0}(\theta_{n})\ldots
a_{M-1}(\theta_{n})]^{T}$ given by
\begin{equation}
\setlength{\abovedisplayskip}{4pt}
\setlength{\belowdisplayskip}{4pt}
\bm{\mathcal{A}}_{n}(i)=
\left(
  \begin{smallmatrix}
    a_{0}(\theta_{n}) & a_{1}(\theta_{n}) & \ldots & a_{I-1}(\theta_{n}) \\
    \vdots & \vdots & \ldots & \vdots \\
    a_{M-I}(\theta_{n}) & a_{M-I+1}(\theta_{n}) & \ldots & a_{M-1}(\theta_{n}) \\
    a_{M-I+1}(\theta_{n}) & a_{M-I+2}(\theta_{n}) & \ldots & 0 \\
    \vdots & \vdots & \ddots & \vdots \\
    a_{M-2}(\theta_{n}) & a_{M-1}(\theta_{n}) & 0 & 0\\
    a_{M-1}(\theta_{n}) & 0 & 0 & 0 \\
  \end{smallmatrix}
\right).
\label{eq:An_matrix}
\end{equation}
The optimization problem in (\ref{eq:cost_function}) can be solved by the
method of Lagrange multipliers whose Lagrangian is described by
\begin{equation}
\setlength{\abovedisplayskip}{1pt}
\setlength{\belowdisplayskip}{1pt}
\begin{split}
\mathcal{L}(i) & =\sum_{l=1}^{i}\alpha^{i-l}
\big|\bm{\bar{\omega}}^{H}(i)\sum_{d=1}^{D}{\bf{q}}_{d}{\bf{d}}^{H}_{d}\bm{\mathcal{R}}(l){\bf{s}}_{d}^{*}(i)\big|^{2} \\
& \quad +2\mathfrak{R}\{\lambda
[\bm{\bar{\omega}}^{H}(i)\sum_{d=1}^{D}{\bf{q}}_{d}{\mathbf{d}}^{H}_{d}\bm{\mathcal{A}}_{n}{\bf{s}}_{d}^{*}(i)-1]\}.
\end{split}
\label{eq:lagrange_cost}
\end{equation}
where $\mathfrak{R}\{\cdot\}$ selects the real part of the argument.
By taking the gradient of (\ref{eq:lagrange_cost}) with respect to ${\bf{s}}_{d}(i)$, we obtain
\begin{equation}
\setlength{\abovedisplayskip}{1pt}
\setlength{\belowdisplayskip}{1pt}
\begin{split}
& \frac{\partial \mathcal{L}(i)}{\partial {\bf{s}}_{d}(i)}  = \sum_{l=1}^{i}\alpha^{i-l} \bm{\mathcal{R}}^{H}(l){\bf{d}}_{d}{\bf{q}}^{H}_{d}\bm{\bar{\omega}}(i)\bm{\bar{\omega}}^{H}(i){\bf{q}}_{d}{\bf{d}}^{H}_{d}\bm{\mathcal{R}}(l){\bf{s}}_{d}^{*}(i)\\
& \quad + \sum_{l=1}^{i}\alpha^{i-l} \bm{\mathcal{R}}^{H}(i){\bf{d}}_{d}{\bf{q}}^{H}_{d}\bm{\bar{\omega}}(i)\bm{\bar{\omega}}^{H}(i)\sum_{j\neq d}^{D}{\bf{q}}_{j}{\bf{d}}^{H}_{j}\bm{\mathcal{R}}(l){\bf{s}}_{j}^{*}(i)\\
& \quad + \lambda^{*} \bm{\mathcal{A}}^{H}_{n}
{\bf{d}}_{d}{\bf{q}}^{H}_{d}\bm{\bar{\omega}}(i).
\end{split}
\label{eq:gradient1}
\end{equation}
By equating (\ref{eq:gradient1}) to zero and solving for ${\bf{s}}_{d}(i)$, we
have
\begin{equation}
\setlength{\abovedisplayskip}{1pt}
\setlength{\belowdisplayskip}{1pt}
{\bf{R}}_{s,d}(i){\bf{s}}_{d}(i)=-\sum_{j\neq d}^{D}{\bf{P}}_{s,j}(i)-\lambda\bm{\mathcal{A}}^{T}_{n}
{\bf{d}}_{d}{\bf{q}}^{H}_{d}\bm{\bar{\omega}}^{*}(i)
\label{eq:normal_eq}
\end{equation}
where
\begin{equation}
\small
\setlength{\abovedisplayskip}{1pt}
\setlength{\belowdisplayskip}{1pt}
{\bf{R}}_{s,d}(i)=\sum_{l=1}^{i}\alpha^{i-l}\bm{\mathcal{R}}^{T}(l){\bf{q}}_{d}{\bf{d}}^{H}_{d}
\bm{\bar{\omega}}^{*}(i)\bm{\bar{\omega}}^{T}(i){\bf{q}}_{d}{\bf{d}}^{H}_{d}\bm{\mathcal{R}}^{*}(l)
\label{eq:R_def}
\end{equation}
\begin{equation}
\setlength{\abovedisplayskip}{1pt}
\setlength{\belowdisplayskip}{1pt}
{\bf{P}}_{s,j}(i)=\sum_{l=1}^{i}\alpha^{i-l}\bm{\mathcal{R}}^{T}(l){\bf{d}}_{d}{\bf{q}}^{H}_{d}
\bm{\bar{\omega}}^{*}(i)\bm{\bar{\omega}}^{T}(i){\bf{q}}_{j}{\bf{d}}^{H}_{j}\bm{\mathcal{R}}^{*}(l){\bf{s}}_{j}(i).
\label{eq:P_def}
\end{equation}
Then $I\times 1$ basis vector ${\bf{s}}_{d}(i)$ is described as:
\begin{equation}
\setlength{\abovedisplayskip}{1pt}
\setlength{\belowdisplayskip}{1pt}
{\bf{s}}_{d}(i)=-{\bf{R}}_{s,d}^{-1}(i)\sum_{j\neq d}^{D}{\bf{P}}_{s,j}(i)-\lambda{\bf{R}}_{s,d}^{-1}(i)\bm{\mathcal{A}}^{T}_{n}{\bf{d}}_{d}{\bf{q}}^{H}_{d}\bm{\bar{\omega}}^{*}(i).
\label{eq:s_d}
\end{equation}
Substituting (\ref{eq:s_d}) into (\ref{eq:cost_function}), we obtain the
Lagrange multiplier:
{\small
\begin{equation} \lambda=\frac{\sum\limits_{j\neq
d}^{D}{\bf{q}}^{H}_{j}\bm{\bar{\omega}}(i){\bf{d}}^{H}_{d}\bm{\mathcal{A}}^{*}_{n}{\bf{s}}_{j}(i-1)-1-
\mathbf{\bm{\prod}}(i){\bf{R}}_{s,d}^{-1}(i)\sum\limits_{j\neq
d}^{D}{\bf{P}}_{s,j}(i)} {\bm{\prod}(i){\bf{R}}_{s,d}^{-1}(i)
\bm{\prod}^{H}(i)} \label{eq:lambda}
\end{equation}
}
where
$\bm{\prod}(i)={\bf{q}}^{H}_{d}\bm{\bar{\omega}}(i){\bf{d}}^{H}_{d}\bm{\mathcal{A}}^{*}_{n}$.
Based on (\ref{eq:s_d}) and (\ref{eq:lambda}), we obtain the $d$th basis vector
${\bf{s}}_{d}(i)$.

Next we consider the update of ${\bf{R}}_{s,d}^{-1}(i)$. By applying the matrix
inversion lemma \cite{Adaptive_Filter_Theory} to (\ref{eq:R_def}), we obtain
\begin{equation}
\begin{split}
\small
{\bf{g}}_{s,d}(i)=\frac
{{\bf{R}}_{s,d}^{-1}(i-1)\bm{\mathcal{R}}^{T}(l){\bf{d}}_{d}}
{\alpha\beta
+{\bf{d}}^{H}_{d}\bm{\mathcal{R}}^{*}(l){\bf{R}}_{s,d}^{-1}(i-1)\bm{\mathcal{R}}^{T}(l){\bf{d}}_{d}}
\end{split}
\label{eq:k_d}
\end{equation}
\begin{equation}
{\bf{R}}_{s,d}^{-1}(i)=\alpha^{-1}{\bf{R}}_{s,d}^{-1}(i-1)-\alpha^{-1}{\bf{g}}_{s,d}(i){\bf{d}}^{H}_{d}\bm{\mathcal{R}}^{*}(l){\bf{R}}_{s,d}^{-1}(i-1)
\label{eq:R_inv_d}
\end{equation}
where
$\beta=({\bf{q}}^{H}_{d}\bm{\bar{\omega}}^{*}(i)\bm{\bar{\omega}}^{T}(i){\bf{q}}_{d})^{-1}$.
As with ${\bf{P}}_{s,j}(i)$, we obtain it through iterations:
\begin{equation}
\setlength{\abovedisplayskip}{5pt}
\setlength{\belowdisplayskip}{5pt}
\begin{split}
{\bf{P}}_{s,j}(i)= & \bm{\mathcal{R}}^{T}(l){\bf{d}}_{d}{\bf{q}}^{H}_{d}\bm{\bar{\omega}}^{*}(i)\bm{\bar{\omega}}^{T}(i)
{\bf{q}}_{j}{\bf{d}}^{H}_{j}\bm{\mathcal{R}}^{*}(l){\bf{s}}_{j}(i)  \\
& + \alpha{\bf{P}}_{s,j}(i-1)
\end{split}
\label{eq:P_j}.
\end{equation}
By employing (\ref{eq:s_d})-(\ref{eq:P_j}), we can update ${\bf{s}}_{d}(i)$ for
$d \in \{1,\ldots D\}$. Given the values of ${\bf{s}}_{d}(i)$, we can compute
$\bm{\bar{\omega}}(i)$. Defining
$\bar{\bf{a}}(i)=\sum_{d=1}^{D}{\bf{q}}_{d}{\mathbf{d}}^{H}_{d}\bm{\mathcal{A}}_{n}{\bf{s}}_{d}^{*}(i)$, (\ref{eq:cost_function}) can be modified as
\begin{equation}
\begin{split}
& \min_{\bm{\bar{\omega}}(i)} \quad \sum_{l=1}^{i}\alpha^{i-l}
\big|\bm{\bar{\omega}}^{H}(i){\bf{\bar{r}}}_{D}(l)\big|^{2}\\
& \textrm{subject to} \quad \bm{\bar{\omega}}^{H}(i)\bar{\bf{a}}(i)=1
\end{split}.
\label{eq:costfunction2}
\end{equation}
Solving for $\bm{\bar{\omega}}(i)$, we have
\begin{equation}
{\bm{g}}_{D}(i)=\frac
{{\bf{R}}_{D}^{-1}(i-1){\bf{\bar{r}}}_{D}(i)}
{\alpha+{\bf{\bar{r}}}^{H}_{D}(i){\bf{R}}_{D}^{-1}(i-1){\bf{\bar{r}}}_{D}(i)}
\label{eq:k_D}
\end{equation}
\begin{equation}
\setlength{\abovedisplayskip}{2pt}
\setlength{\belowdisplayskip}{2pt}
{\bf{R}}_{D}^{-1}(i)=\alpha^{-1}{\bf{R}}_{D}^{-1}(i-1)
-\alpha^{-1}{\bm{g}}_{D}(i){\bf{\bar{r}}}^{H}_{D}(i){\bf{R}}_{D}^{-1}(i-1)
\label{eq:R_inv_D}
\end{equation}
\begin{equation}
\setlength{\abovedisplayskip}{2pt}
\setlength{\belowdisplayskip}{2pt}
\bm{\bar{\omega}}(i)=\frac
{{\bf{R}}_{D}^{-1}(i)\bar{\bf{a}}(i)}
{\bar{\bf{a}}^{H}(i){\bf{R}}_{D}^{-1}(i)\bar{\bf{a}}(i)}
\label{eq:w_D}
\end{equation}
where
${\bf{R}}_{D}^{-1}(i)=\sum_{l=1}^{i}\alpha^{i-l}{\bf{\bar{r}}}_{D}(i){\bf{\bar{r}}}^{H}_{D}(i)$.
Based on the previous derivations, we calculate the output power for each
scanning angle $\theta_{n}$:
\begin{equation}
P(\theta_{n})=\sum_{l=1}^{i}\alpha^{i-l}
\big|\bm{\bar{\omega}}^{H}(i){\bf{\bar{r}}}_{D}(i)\big|^{2}=\frac
{1}
{\bar{\bf{a}}^{H}(i){\bf{R}}_{D}^{-1}(i)\bar{\bf{a}}(i)}
\label{eq:Power}.
\end{equation}
The ALRD-RLS algorithm is summarized in Table
\ref{tab:ALRD-RLS}. 

\begin{table}[hh]
\centering
\caption{The ALRD-RLS DOA estimation algorithm.}
\begin{tabular}{ll} \hline
1~~~Set $N$ and $\alpha$ \\
2~~~for each scanning angle $\theta_{n}$ do  \\
3~~~~~~Initialize $\mathbf{R}_{s,d}^{-1}(0)$, $\mathbf{P}_{s,j}(0)$, ${\bf{s}}_{d}(0)$, $\bm{\bar{\omega}}(0)$\\
4~~~~~~for each snapshot $i$ ($i=1,\ldots,N$) do \\
5~~~~~~~~~ for each basis $d$ ($d=1, \ldots, D$) do \\
6~~~~~~~~~~~~ Update ${\bf{s}}_{d}(i)$ based on (\ref{eq:s_d})-(\ref{eq:P_j}).\\
7~~~~~~~~~ Update $\bar{\bm{\omega}}(i)$ based on (\ref{eq:k_D})-(\ref{eq:w_D}).\\
8~~~~~~Calculate the output power $P(\theta_{n})=(\bar{\bf{a}}^{H}(N){\bf{R}}_{D}^{-1}(N)\bar{\bf{a}}(N))^{-1}$\\
9~~~Determine the estimated DOA $\hat{\bm{\theta}}=\textrm{arg}~\max_{\theta_{n}}~P(\theta_{n})$\\
\hline
\end{tabular}
\label{tab:ALRD-RLS}
\end{table}

\subsection{Proposed MALRD-RLS DOA Estimation Algorithm}

By examining the structure of the ALRD scheme, we can reduce its computational
cost by using a single basis vector in the decomposition matrix. From this
observation, we come up with a modified version of the ALRD-RLS algorithm,
i.e., the MALRD-RLS algorithm. Specifically, the columns of the decomposition
matrix in the MALRD-RLS algorithm are formed by shifted versions of the same
basis vector ${\bf{s}}(i)$, which results in
${\bf{\bar{r}}}_{D}(i)={\bf{Q}}\bm{\mathcal{R}}(i){\bf{s}}^{*}(i)$, where
${\bf{Q}}=\sum_{d=1}^{D}{\bf{q}}_{d}{\mathbf{d}}^{H}_{d}$. Therefore, the
optimization problem solved by the MALRD-RLS algorithm is:
\begin{equation}
\setlength{\abovedisplayskip}{2pt}
\setlength{\belowdisplayskip}{2pt}
\begin{split}
& \min_{\bm{\bar{\omega}}(i),\mathbf{s}(i)} \sum_{l=1}^{i}\alpha^{i-l}
\big|\bm{\bar{\omega}}^{H}(i){\bf{Q}}\bm{\mathcal{R}}(l){\bf{s}}^{*}(i)\big|^{2}\\
& {\rm subject~~ to} \quad
\bm{\bar{\omega}}^{H}(i){\bf{Q}}\bm{\mathcal{A}}_{n}{\bf{s}}^{*}(i)=1
\end{split}.
\label{eq:cost_function2}
\end{equation}
This problem can be solved by following the same procedure as in the ALRD-RLS
algorithm. Firstly, we construct the Lagrangian function as
\begin{equation}
\setlength{\abovedisplayskip}{2pt}
\setlength{\belowdisplayskip}{2pt}
\begin{split}
\mathcal{L}(i) & =\sum_{l=1}^{i}\alpha^{i-l}
\big|\bm{\bar{\omega}}^{H}(i){\bf{Q}}\bm{\mathcal{R}}(l){\bf{s}}^{*}(i)\big|^{2} \\
& +2\mathfrak{R}\{\lambda
[\bm{\bar{\omega}}^{H}(i){\bf{Q}}\bm{\mathcal{A}}_{n}{\bf{s}}^{*}(i)-1]\}.
\end{split}
\label{eq:lagrange_cost_malrd}
\end{equation}
Secondly, we take the gradient of (\ref{eq:lagrange_cost_malrd}) with respect
to ${\bf{s}}(i)$, set the result to zero and solve for ${\bf{s}}(i)$. The
update equation of ${\bf{s}}(i)$ is given by
\begin{equation}
\setlength{\abovedisplayskip}{2pt}
\setlength{\belowdisplayskip}{2pt}
{\bf{s}}(i)=\frac
{{\bf{R}}_{s}^{-1}(i)\bm{\mathcal{A}}_{n}^{T}{\bf{Q}}^{T}\bm{\bar{\omega}}^{*}(i)}
{\bm{\bar{\omega}}^{T}(i){\bf{Q}}\bm{\mathcal{A}}_{n}^{*}{\bf{R}}_{s}^{-1}(i)\bm{\mathcal{A}}_{n}^{T}{\bf{Q}}^{T}\bm{\bar{\omega}}^{*}(i)}
\label{eq:s}
\end{equation}
where
${\bf{R}}_{s}(i)=\sum_{l=1}^{i}\alpha^{i-l}\bm{\mathcal{R}}^{T}(l){\bf{Q}}^{T}
\bm{\bar{\omega}}^{*}(i)\bm{\bar{\omega}}^{T}(i){\bf{Q}}\bm{\mathcal{R}}^{*}(l)$.
The matrix ${\bf{R}}_{s}^{-1}(i)$ can be computed as:
\begin{equation}
{\bf{g}}_{s}(i)=\frac
{{\bf{R}}_{s}^{-1}(i-1)\bm{\mathcal{R}}^{T}(i){\bf{Q}}^{T}
\bm{\bar{\omega}}^{*}(i)}
{\alpha+\bm{\bar{\omega}}^{T}(i){\bf{Q}}\bm{\mathcal{R}}^{*}(i){\bf{R}}_{s}^{-1}(i-1)\bm{\mathcal{R}}^{T}(i){\bf{Q}}^{T}
\bm{\bar{\omega}}^{*}(i)}
\label{eq:k_d_malrd}
\end{equation}
\begin{equation}
\small {\bf{R}}_{s}^{-1}(i) = \alpha^{-1}{\bf{R}}_{s}^{-1}(i-1)
-\alpha^{-1}{\bf{g}}_{s}(i)\bm{\bar{\omega}}^{T}(i){\bf{Q}}\bm{\mathcal{R}}^{*}(i){\bf{R}}_{s}^{-1}(i-1).
\label{eq:R_inv_malrd}
\end{equation}
Next, we discuss the update of $\bm{\bar{\omega}}(i)$. By redefining
$\bar{\bf{a}}(i)={\bf{Q}}\bm{\mathcal{A}}_{n}{\bf{s}}^{*}(i)$, the cost
function for the update of $\bm{\bar{\omega}}(i)$ is the same as that in
(\ref{eq:costfunction2}). Hence $\bm{\bar{\omega}}(i)$ can also be constructed
by (\ref{eq:k_D})-(\ref{eq:w_D}) in the MALRD-RLS algorithm.

\begin{table}[hh]
\centering
\caption{The MALRD-RLS DOA estimation algorithm.}
\begin{tabular}{ll} \hline
1~~~Set $N$ and $\alpha$ \\
2~~~for each scanning angle $\theta_{n}$ do  \\
3~~~~~~Initialize $\mathbf{R}_{s}^{-1}(0)$, ${\bf{s}}(0)$, $\bm{\bar{\omega}}(0)$\\
4~~~~~~for each snapshot $i$ ($i=1,\ldots,N$) do \\
5~~~~~~~~~ Update ${\bf{s}}(i)$ based on (\ref{eq:s})-(\ref{eq:R_inv_malrd}).\\
6~~~~~~~~~ Redefine $\bar{\bf{a}}(i)={\bf{Q}}\bm{\mathcal{A}}_{n}{\bf{s}}^{*}(i)$ and update $\bar{\bm{\omega}}(i)$ based on (\ref{eq:k_D})-(\ref{eq:w_D}).\\
7~~~~~Calculate the output power $P(\theta_{n})=(\bar{\bf{a}}^{H}(N){\bf{R}}_{D}^{-1}(N)\bar{\bf{a}}(N))^{-1}$\\
8~~~Determine the estimated DOA $\hat{\bm{\theta}}=\textrm{arg}~\max_{\theta_{n}}~P(\theta_{n})$\\
\hline
\end{tabular}
\label{tab:MALRD-RLS}
\end{table}

A brief summary of the MALRD-RLS algorithm is illustrated in Table
\ref{tab:MALRD-RLS}. After the update of ${\bf{s}}(i)$ and
$\bar{\bm{\omega}}(i)$, we calculate the output power spectrum based on
$P(\theta_{n})=\sum_{l=1}^{i}\alpha^{i-l}
\big|\bm{\bar{\omega}}^{H}(i){\bf{\bar{r}}}_{D}(i)\big|^{2}=
(\bar{\bf{a}}^{H}(i){\bf{R}}_{D}^{-1}(i)\bar{\bf{a}}(i))^{-1}$. The peaks of
the power spectrum are the estimated DOAs.

\subsection{Computational Complexity}
\label{ssec:Computational_Complexity}
Here we detail the computational complexity of the proposed ALRD-RLS and
MALRD-RLS algorithms and several existing DOA estimation algorithms. ESPRIT
uses an EVD of $\mathbf{R}$, which has complexity of $O(M^{3})$. MUSIC employs
both the EVD and grid search, resulting in a cost of
$O(M^{3}+(180/\triangle)M^{2})$, with $\triangle$ being the search step. Matrix inversions and grid searches are
essential for Capon, whose complexity is $O((180/\triangle)M^{3})$. For the AV
and CG algorithms, the construction of the basis vectors leads to a complexity
which is higher than that of the ESPRIT algorithm \cite{4063549}\cite{CG}. The
JIO-RLS algorithm has a cost of $O((180/\triangle)N(M^{2}+D^{2}))$, with $D$
being the length of the reduced-rank received vector.
ALRD-RLS avoids the EVD,
the matrix inversion and the construction of the transformation matrix, and the
update of $D$ basis vectors and an auxiliary parameter vector requires
$O((180/\triangle)N(DI^{2}+D^{2}))$. MALRD-RLS only uses one basis vector and
costs $O((180/\triangle)N(I^{2}+D^{2}))$. The computational complexity of the
analyzed algorithms is depicted in Table \ref{tab:complexity}. Even in a large
sensor array, $I$ and $D$ are small numbers, with $I\ll M$ and $D\ll M$, the
cost of ALRD-RLS and MALRD-RLS can be less than those of the existing algorithms.
\begin{table}[h]
\centering
\caption{Comparison of Computational Complexity.}
{
\vspace{-0.3em}
\begin{tabular}{cc}
\hline
\bf Algorithm & \bf Complexity \\
\hline
ESPRIT \cite{32276}  & {$O(M^{3})$}   \\
MUSIC \cite{1143830}  & {$O(M^{3}+(180/\triangle)M^{2})$}   \\
Capon \cite{1449208}  & {$O((180/\triangle)M^{3})$}   \\
AV \cite{4063549}  &  {$O((180/\triangle)KM^{2})$}  \\
CG \cite{CG}  & {$O((180/\triangle)KM^{2})$}   \\
JIO-RLS \cite{6978860}  & {$O((180/\triangle)N(M^{2}+D^{2})$}   \\
ALRD-RLS  & {$O((180/\triangle)N(DI^{2}+D^{2})$}   \\
MALRD-RLS  & {$O((180/\triangle)N(I^{2}+D^{2})$}   \\
\hline
\label{tab:complexity}
\end{tabular}
}
\end{table}

\vspace{-0.8em}
\section{Simulations}
\label{sec:simulation}

In this section, we evaluate the ALRD-RLS and MALRD-RLS algorithms
through simulations. We compare ALRD-RLS and MALRD-RLS with MUSIC,
ESPRIT, Capon, CG, AV and the JIO-RLS algorithms. A ULA with $M=60$
elements is adopted in the experiments. $K=15$ narrowband source
signals impinge on the ULA from directions
$\theta_{1},\ldots,\theta_{K}$, with $2$ of them being correlated
and the others uncorrelated. The correlated source samples are
generated from a first-order autoregressive process:
\begin{equation}
\setlength{\abovedisplayskip}{3pt}
\setlength{\belowdisplayskip}{3pt}
s_{1}\sim\mathcal{N}(0,\sigma^{2}_{s})~~\textrm{and}~~s_{2}(i)=\varrho s_{1}(i)+\sqrt{1-\varrho^{2}}e(i)
\label{eq:correlated_signals}
\end{equation}
where $e\sim\mathcal{N}(0,\sigma^{2}_{s})$. $\varrho$ is the correlation
coefficient fixed as $0.7$ in this work. We assume that a small
number of snapshots are available for DOA estimation and fix $N=20$ in the
simulations. The source signals with powers $\sigma^{2}_{s}=1$ are modulated by the binary phase
shift keying (BPSK) scheme. The search step is chosen as $0.3^{o}$ for the
algorithms based on grid search. We assume that the source DOAs are resolved if
$|\hat{\theta}_{k}-\theta_{k}|<|\theta_{k}-\theta_{k-1}|/2$. In each
experiment, 100 independent Monte Carlo runs are conducted to obtain the
curves.

\begin{figure}[htbp]
\centering\scalebox{0.5}{\includegraphics{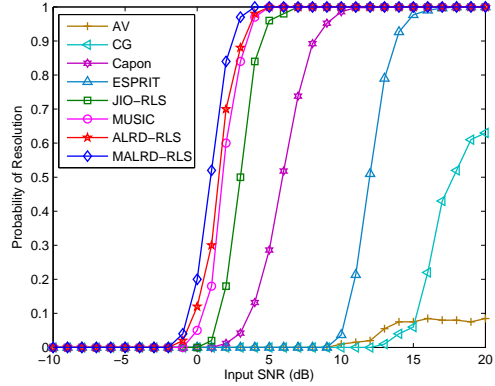}} \vspace{-0.5em}
\caption{\small Probability of resolution versus input SNR.}
\label{fig:mixed_por}
\end{figure}

In the first example shown in Fig. \ref{fig:mixed_por}, we plot the
resolution probability versus the input signal-to-noise ratio (SNR)
of the analyzed algorithms. We set the parameters for JIO-RLS to
$D=5$ and $\lambda=0.998$. For MALRD-RLS and ALRD-RLS, we choose
$I=12$, $D=5$, $\lambda=0.998$. Note that higher $I$ and $D$ yield
higher probability of resolution, yet they lead to higher cost as
well. We have examined the values of $I, D\in[3,15]$ and observed
that $I=12$, $D=5$ provides a satisfactory performance with
acceptable complexity. MALRD-RLS achieves the best performance in
the large sensor arrays for different SNR values, followed by
ALRD-RLS, MUSIC, JIO-RLS, Capon and ESPRIT. The AV and CG algorithms
fail to resolve the DOAs for most of the SNR values when many
sources are present. Note that MUSIC, ESPRIT, Capon, AV and CG
require forward backward averaging (FBA) \cite{fba,301848} to ensure
satisfactory performance for correlated signals.

\begin{figure}[htbp]
\centering\scalebox{0.5}{\includegraphics{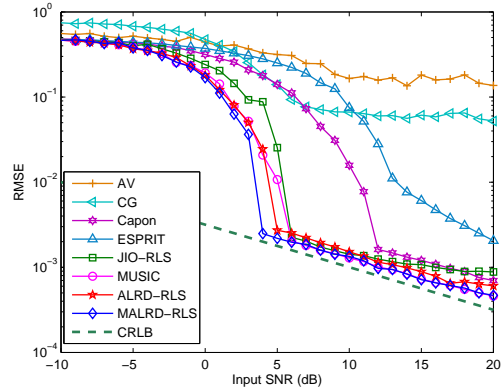}} \vspace{-0.5em}
\caption{\small RMSE versus input SNR.} \label{fig:mixed_RMSE}
\end{figure}

We then evaluate the root mean square error (RMSE) performance of the analyzed
algorithms, which is calculated as
$RMSE=\sqrt{\frac{1}{NK}\sum_{n=1}^{N}\sum_{k=1}^{K}(\hat{\theta}_{k}-\theta_{k})^{2}}$.
From Fig. \ref{fig:mixed_RMSE}, MALRD-RLS provides a superior RMSE performance
with the lowest threshold SNR and the lowest RMSE level in high SNRs. The gap
between the RMSE of MALRD-RLS and the Cramer-Rao bound (CRB) is due to the
small number of available snapshots and the fact that the correlated sources
degrade the performance.
\vspace{-1em}

\vspace{-1em}
\section{Conclusion}
\label{sec:conclusion} In this paper, we have proposed the ALRD scheme and the
ALRD-RLS and MALRD-RLS subspace DOA estimation algorithms based on alternating
optimization. The proposed algorithms are suitable for large sensor arrays and
have a lower computational cost than existing techniques. Simulation results
show that MALRD-RLS and ALRD-RLS  outperform previously reported algorithms.

\bibliographystyle{IEEEtran}
\bibliography{reference}

\end{document}